\documentclass[12pt,multibox]{article}
\usepackage{amsthm,latexsym,multibox,amssymb,amsfonts,amsbsy}
\usepackage{graphicx,lscape,fancyhdr,array}
\usepackage{cite}
 \newcommand{\ft}{\footnotesize}
 \def\a{\alpha}
 \def\b{\beta}
 \def\g{\gamma}
 \def\d{\delta}
 
 \def\h{\eta}
 \def\l{\lambda}

 \def\m{\mu}
 \def\n{\nu}
 \def\r{\rho}
 \def\o{\omega}
 \def\s{\sigma}
 \def\S{\Sigma}

 \def\pa{\partial}

 \def\be{\begin{equation}}
 \def\ee{\end{equation}}
 \def\beq{\begin{eqnarray}}
 \def\eeq{\end{eqnarray}}

 \newcommand{\bqn}{\begin{eqnarray}}\newcommand{\eqn}{\end{eqnarray}}
 \newtheorem{theorem}{Theorem}[section]
 \newtheorem{lemma}{Lemma}[section]

\begin{document}

 \begin{titlepage}
 \begin{flushright}
ULB-TH/04-04 \\
DAMTP-2004-27
 \end{flushright}
 \vskip 2cm

 \begin{centering}

 {\large {\bf Consistent deformations of $[p,p]$--type gauge field theories }}

 \vspace{2cm}
 Nicolas Boulanger$^{a}$ and
 Sandrine Cnockaert$^{b,}\footnote{``Aspirant du F.N.R.S., Belgium''}$ \\
 \vspace{1.5cm}
 {\small
 $^a$ Department of Applied Mathematics and Theoretical Physics\\ 
 Wilberforce Road, Cambridge CB3 0WA, UK\\
 \vspace{.2cm}
 $^b$ Physique Th\'eorique et Math\'ematique and
International Solvay Institutes,\\
 Universit\'e Libre
 de Bruxelles,  C.P. 231, B-1050, Bruxelles, Belgium }     \\
 
 \vspace{.7cm}
{\tt N.Boulanger@damtp.cam.ac.uk, Sandrine.Cnockaert@ulb.ac.be}

 \vspace{1.5cm}
 
 \end{centering}

 \begin{abstract}
Using BRST-cohomological techniques, we analyze the consistent deformations of theories 
describing free  tensor gauge fields whose symmetries are represented by Young tableaux 
 made of two columns of equal length $p\,$, $p>1\,$. Under the assumptions of locality and 
Poincar\'e invariance, we find that there is no consistent deformation of these theories that 
non-trivially modifies the gauge algebra and/or the gauge transformations. Adding the requirement 
that the deformation contains no more than two derivatives, the only possible deformation is a 
cosmological-constant-like term.   
 \end{abstract}

 \vfill
 \end{titlepage}

 \section{Introduction}

Recently, there has been a surge of interest in the study of theories for gauge fields 
transforming in ``exotic" irreducible representations of the general linear group $GL(D)\,$ 
\cite{Mirian,CMR,DVH2,BB1,dMH1,BBH,BB2,dMH2,BCH,Zinoviev,Vasi+dM,Biz}.
Such gauge fields are said to have mixed symmetries since they are neither completely 
symmetric nor completely antisymmetric.
They are believed to play an important role in the description of tensionless string theories, 
where the higher-spin excitation modes become all massless 
\cite{Bon, MirianSagnotti} (note that, in the context of string field theories, the ``exotic" 
higher spin fields naturally appear in \textsl{reducible} representations of $GL(D)\,$).

In this paper we investigate the possible consistent deformations of free theories 
whose gauge fields $C_{\m_1\ldots\m_p\vert\n_1\ldots\n_p}$ possess the Young symmetry type\footnote{An irreducible representation of the general linear group $GL(D)$ is denoted by $[c_1,c_2,\ldots,c_L]\,$, where $c_i$ indicates the number of boxes in the $i$th column of the Young tableau characterizing the corresponding irreducible representation. 
A $p$-form gauge field is then denoted by $[p]$ whereas a rank-$s$ 
symmetric tensor is denoted by $[\;\underbrace{1,1,\ldots,1}_{\mbox{\tiny{s times}}}\;]\,$.
\label{conventions}} $[p,p]\,$, with $p>1$ but otherwise arbitrary:
\begin{center} 
\begin{picture}(0,40)(-20,-23)
\put(-100,-8){$C_{\m_1\ldots\m_p\vert\n_1\ldots\n_p}~~\sim$}
\multiframe(0,10)(10.5,0){1}(10,10){\ft$1$}
\multiframe(10.5,10)(10.5,0){1}(10,10){\ft$1$}
\multiframe(0,-0.5)(10.5,0){1}(10,10){\ft$2$}
\multiframe(10.5,-0.5)(10.5,0){1}(10,10){\ft$2$}
\multiframe(0,-18)(10.5,0){1}(10,17){$ $} \put(4,-13.5){$\vdots$}
\multiframe(10.5,-18)(10.5,0){1}(10,17){$ $} \put(14.5,-13.5){$\vdots$}
\multiframe(0,-28.5)(10.5,0){1}(10,10){\ft$p$}
\multiframe(10.5,-28.5)(10.5,0){1}(10,10){\ft$p$}
\put(34.5,-8){$.$}
\end{picture}
\end{center}
The case $p=1$ corresponds to linearized gravity and is treated in Ref. \cite{BDGH}. 
We determine all the consistent, local interactions that these free theories possibly 
admit.
By ``consistent'' we mean that the deformation is smooth in a formal deformation 
parameter $g$ and reduces to the original theory in the free limit where $g=0\,$.

Our approach is based on the BRST-cohomology-antifield formalism \cite{BH93}, 
in which consistent 
deformations of the action appear as deformations of the solutions of the master equation 
that preserve the master equation. The first-order consistent deformations define 
cohomological classes of the BRST differential at ghost number zero. 
We will not devote much time to the explanation of this technique; for more details and 
references, see e.g. \cite{BDGH} where the same BRST-antifield tools are used in the context of linearized Einstein theory.

Our result is that the free theory for the exotic field 
$C_{\m_1\ldots\m_p \vert \n_1\ldots\n_p}$ admits no consistent local deformation that  
is compatible with Poincar\'e invariance and that non-trivially modifies the gauge 
algebra and/or the gauge transformations. 
This result holds without any assumption on the number of derivatives in the 
expressions for the gauge transformations and the interaction terms. 
Moreover, it is not required that the deformed gauge symmetries should close off-shell. 
If we demand that the deformations of the free Lagrangian 
should not contain more than two derivatives of the fields (i.e., the allowed interaction 
terms under consideration schematically read ``$C\cdots C$", ``$C\cdots C\cdot\partial C$", ``$C\cdots C 
\cdot \partial^2 C$", or ``$C\cdots C \cdot\partial C\cdot \partial C$"), then 
the only possible deformation is a cosmological-constant-like term $a_0= \Lambda C^{[p]}\,$, where 
$C^{[p]}$ denotes the $p$--th trace of the field $C_{\m_1\ldots\m_p \vert 
\n_1\ldots\n_p}\,$.  

 \setcounter{equation}{0} \setcounter{theorem}{0}
 \setcounter{lemma}{0} 

 \section{The free theory}
 \setcounter{equation}{0} \setcounter{theorem}{0}

 We work in arbitrary spacetime dimension D, only taken to be strictly greater than $2p+1$
, so that the theory describes local degrees of freedom. \\

The free Lagrangian for the gauge field $C_{\m_1\ldots\m_p\vert\n_1\ldots\n_p}$ reads
 \be {\cal
 L}_0=(-)^{p}\frac{(p+1)}{2}\;\delta^{[\r_1 \dots \r_p \m_1 \dots \m_{p+1}]}_{[\n_1 \dots \n_p \s_1 \dots \s_{p+1}]}\; T_{\s_1 \dots \s_{p+1}\vert \r_1 \dots \r_p}\;  T_{\m_1 \dots \m_{p+1}\vert\n_1 \dots \n_p }\,, \label{Lagrangian}
 \ee 
 where $T$ is the tensor \be  T_{\m_1 \dots \m_{p+1}\vert\n_1 \dots \n_p }=\pa_{[\m_1} C_{\m_2 \dots \m_{p+1}]\vert\n_1 \dots \n_p }  ,\ee
the square brackets $[\ldots ] $ denote antisymmetrization with global weight one and $\delta^{\m_1 \ldots \m_n}_{\n_1 \ldots \n_n} \equiv \delta^{\m_1 }_{\n_1 } \ldots \delta^{\m_n }_{\n_n }$.
 
 The gauge transformations are 
 \be
 \delta_{\a} C_{\m_1 \dots \m_p \vert \n_1 \dots \n_p}=
\a_{\m_1 \dots \m_p \vert [\n_1 \dots \n_{p-1},\n_p]}+\a_{\n_1 \dots \n_p \vert [\m_1 \dots \m_{p-1},\m_p]}\,,
 \label{invdejauge}\\
 \ee 
 where $\a_{,\n} \equiv \pa_{\n}\a$ and $\a_{\m_1 \dots \m_p \vert \n_1 \dots \n_{p-1}}$ is an arbitrary $[p,p-1]$-tensor field, that thus possesses the following algebraic properties:
\begin{eqnarray}
&\a_{\m_1 \dots \m_p \vert \n_1 \dots \n_{p-1}}=\a_{[\m_1 \dots \m_p ] \vert \n_1 \dots \n_{p-1}}=\a_{\m_1 \dots \m_p \vert [\n_1 \dots \n_{p-1}]}\,,&
\nonumber \\
&\a_{[\m_1 \dots \m_p \vert \m_{p+1} ] \n_2 \dots \n_{p-1}}=0\,.&
\end{eqnarray}

 To obtain a  gauge-invariant object, one must take two derivatives of the field. The tensor  
 \be
K_{\m_1 \dots \m_{p+1} \vert \n_1 \dots \n_{p+1} }=\pa_{[\m_1}C_{\m_2 \dots \m_{p+1} ]\vert [\n_1 \dots \n_p ,\n_{p+1} ]}
 \label{fieldstrength} \ee 
 is easily verified to be gauge
 invariant; moreover its vanishing implies that $C_{\m_1 \dots \m_p \vert \n_1 \dots \n_p}$ is pure-gauge \cite{BB1}.
 The most general gauge invariant object depends on the
 field $C_{\m_1 \dots \m_p \vert \n_1 \dots \n_p}$ and its derivatives only through the
 ``curvature" $K_{\m_1 \dots \m_{p+1} \vert \n_1 \dots \n_{p+1} }$ and its derivatives.  

 The equations of motion read  
 \be 0=G^{\m_1\ldots\m_p\vert}_{\;\;\quad\quad\n_1\ldots\n_p}\equiv(-)^{p+1}(p+1)\;\delta^{[\r_1 \dots \r_{p+1} \m_1 \dots \m_{p}]}_{[\n_1 \dots \n_p \s_1 \dots \s_{p+1}]}\; K_{\s_1 \dots \s_{p+1}\vert \r_1 \dots \r_{p+1}}\,. \label{ELeq}
 \ee 
 They are equivalent to
\begin{eqnarray}
        \h^{\m\n}K_{\m\m_1\ldots\m_p\vert\n\n_1\ldots\n_p} = 0\,.
\end{eqnarray}
Because the action is gauge-invariant, the equations of motion
 satisfy the ``Bianchi identities" 
 \be
 \pa_{\m_1}G^{\m_1\ldots\m_p\vert}_{\;\;\quad\quad\n_1\ldots\n_p}\equiv 0\,.
 \label{Bianchi}
 \ee 
 An easy way to check these identities is to
 observe that one has \be G^{\m_1\ldots\m_p\vert\n_1\ldots\n_p}\equiv
 \pa_{\a\b}M^{\a\m_1 \dots \m_p \vert \b\n_1 \dots \n_q}\,,\ee where the $[p+1,p+1]$-type  tensor $M^{\a\m_1 \dots \m_p \vert \b\n_1 \dots \n_q}$ is given by
 \be
M^{\a\m_1 \dots \m_p \vert \b\n_1 \dots \n_p}= (-)^{pq+1}(p+1)\;\delta^{[\r_1 \dots \r_p \a  \m_1 \dots \m_{p}]}_{[\n_1 \dots \n_p \b \s_1 \dots \s_p]}\; C_{\s_1 \dots \s_p\vert \r_1 \dots \r_p}\,.
\ee

 The gauge symmetries (\ref{invdejauge}) are reducible. Indeed, \be
 \d_{\tilde{\a}^{(1)}}C_{\m_1 \dots \m_p \vert \n_1 \dots \n_p}\equiv 0 \ee when \be
 \tilde{\a}^{(1)}_{\m_1 \dots \m_p \vert \n_1 \dots \n_{p-1}}=\tilde{\a}^{(2)}_{\m_1 \dots \m_p \vert [\n_1 \dots \n_{p-2},\n_{p-1}]} + \frac{p}{2}\;\tilde{\a}^{(2)}_{\n_1 \dots \n_{p-1}[ \m_p \vert \m_1 \dots \m_{p-2},\m_{p-1}]}\,,
\label{reduc} \ee 
where
 $\tilde{\a}^{(2)}_{\m_1 \dots \m_p \vert \n_1 \dots \n_{p-2}}$ is an arbitrary field of Young symmetry type $[p,p-2]\,$. For $p>2$, there is further local reducibility.
There are altogether $p-1$ reducibilities, involving arbitrary $[p,p-i]$-type fields $\tilde{\a}^{(i)}_{\m_1 \dots \m_p \vert \n_1 \dots \n_{p-i}}\,$, $2\leq i \leq p\,$.

\section{Deforming the free theory}
\label{sec:Deformingthefreetheory}

 The problem of introducing (smooth) consistent interactions is
 that of simultaneously deforming the free Lagrangian, the gauge transformations and the reducibility conditions: 
\bqn
 {\cal L} &=& {\cal L}_0 + g {\cal L}_1 + g^2 {\cal L}_2 +
 \cdots\,, \\
 \delta_{\a} C&=& \hbox{(\ref{invdejauge})} + g
 \,\delta_{\a}^{(1)} C + g^2 \, \delta_{\a}^{(2)}
 C+ \cdots \,,\\
\delta_{\a} \tilde{\a}^{(i)}&=& \delta_{\a}^{(0)} \tilde{\a}^{(i)}+ g
 \,\delta_{\a}^{(1)} \tilde{\a}^{(i)} + g^2 \, \delta_{\a}^{(2)}
 \tilde{\a}^{(i)}+ \cdots \,,
\eqn  
 in such a way that the deformed Lagrangian is invariant under the deformed gauge  transformations, the latter obeying the new reducibility conditions.
 
 We shall impose the further requirement that the first order
 vertex ${\cal L}_1$ be Poincar\'e-invariant and local.  Under this sole
 condition (together with consistency), we show that the gauge algebra together with 
 the gauge transformations cannot be deformed.
  
 Adding the requirement that the deformed Lagrangian should not contain more than two derivatives, we further show that, except for a cosmological-constant-like term, there is no deformation that would only modify the free Lagrangian, leaving the gauge structure unchanged. 
 
 \subsection{BRST differential and field content}

 As shown in \cite{BH93}, the first-order consistent local
 interactions correspond to elements of the cohomology $H^{D,0}(s
 \vert d)$ of the BRST differential $s$ modulo the total derivative $d$, 
 in maximum form-degree $D$ and in ghost
 number $0$. That is, one must compute the general solution of the
 cocycle condition \be s a + db =0, \label{coc}
 \ee 
 where $a$ is a
 $D$-form of ghost number zero and $b$ a $(D-1)$-form of ghost
 number one, with the understanding that two solutions $a$ and $a'$
 of (\ref{coc}) that differ by a trivial solution \be a' = a + s m
 + dn \ee 
 should be identified as they define the same interactions
 up to field redefinitions. 
 The cocycles and coboundaries $a,b,m,n,\ldots\,$ are functions of the field variables 
 (including ghosts and antifields) and their derivatives, up to some finite order in the 
 derivatives. They are called ``local functions".
 Given a non-trivial cocycle $a$ of $H^{D,0}(s \vert d)$, the corresponding first-order  interaction vertex ${\cal L}_1$ 
 is obtained by setting the ghosts equal to zero.

  Since the
 theory at hand is a free theory, the BRST differential takes the
 simple form \be s = \delta + \gamma\,, \ee where $\d$ is the Koszul-Tate differential and $\gamma$ is the exterior derivative along the gauge orbits. A grading is associated to
 each of these differentials: $\g$ increases by one unit the
 ``pure ghost number" denoted {\it{puregh}}, while $\d$ decreases
 the ``antighost number'' {\it{antigh}} by one unit. The antighost number is also called ``antifield number''. The ghost
 number {\it{gh}} is defined by \be
 {\it{gh}}={\it{puregh}}-{\it{antigh}}. \ee

Using the property $s^2
 =0$, one concludes that \be \delta^2 = 0\,, \; \delta \gamma +
 \gamma \delta = 0\,, \; \gamma^2 = 0\,. \ee

 In the theories under consideration and according to the general rules of the BRST-antifield formalism, the spectrum of fields (including ghosts) and
 antifields is given by
 \begin{itemize}
 \item the field $C_{\m_1 \dots \m_p \vert \n_1 \dots \n_p}\,$, with ghost number zero and
 antifield number zero; 
 \item the ghosts $A^{(i)}_{\m_1 \dots \m_p\vert \n_1 \dots \n_{p-i}}$ with ghost number $i$ and antifield number zero, where $1 \leq i \leq p\,$;
\item the antifield $C^{*\m_1 \dots \m_p \vert \n_1 \dots \n_p}$, with ghost
 number minus one and antifield number one; 
\item the antifields
 $A^{*(i)\m_1 \dots \m_p\vert \n_1 \dots \n_{p-i}}$ with ghost number minus $(i+1)$ and
 antifield number $(i+1)$, where $1 \leq i \leq p\,$.
 \end{itemize}

 The pureghost number, antighost number, ghost
 number and grassmannian parity of the various fields are displayed in the following table:
 \vspace{5mm}

 \begin{center}
 \begin{tabular}{|c|c|c|c|c|}
 \hline Z  & $puregh(Z)$  & $antigh(Z)$  & $gh(Z)$  & parity (mod $2$)\\ \hline
 $C_{\m_1 \dots \m_p \vert \n_1 \dots \n_p}$   &$0$  & $0$  &$0$ &$0$ \\
 $A^{(i)}_{\m_1 \dots \m_p\vert \n_1 \dots \n_{p-i}}$ & $i$ & $0$ & $i$ & $i$ \\
 $C^{*\m_1 \dots \m_p \vert \n_1 \dots \n_p}$ & $0$& $1$ & $-1$ & $1$ \\
 $A^{*(i)\m_1 \dots \m_p\vert \n_1 \dots \n_{p-i}}$ & $0$ & $i+1$ & $-(i+1)$ & $i+1$ \\
 \hline
 \end{tabular}
 \end{center}
 
 The action of the differentials  $\delta$ and $\gamma$ is zero on all the
 fields of the formalism except in the following cases : 
\begin{eqnarray}
\d C^{*\m_1 \dots \m_p \vert \n_1 \dots \n_p}&=&G^{\m_1 \dots \m_p \vert \n_1 \dots \n_p}
\nonumber \\
\d A^{*(1)\m_1 \dots \m_p\vert \n_1 \dots \n_{p-1}}&=&-2\pa_{\s}C^{*\m_1 \dots \m_p \vert \n_1 \dots \n_{p-1} \s}
\nonumber \\
\d A^{*(i)\m_1 \dots \m_p\vert \n_1 \dots \n_{p-i}}&=&(-)^i \pa_{\s}A^{*(i-1)\m_1 \dots \m_p\vert \n_1 \dots \n_{p-i}\s}\,,\quad 2\leq i \leq p\nonumber        
\end{eqnarray}
and
 \bqn \g
 C_{\m_1 \dots \m_p \vert \n_1 \dots \n_p}&=&A^{(1)}_{\m_1 \dots \m_p \vert [\n_1 \dots \n_{p-1},\n_{p}]} + A^{(1)}_{\n_1 \dots \n_{p}\vert [\m_1 \dots \m_{p-1},\m_{p}]} \nonumber \\
\g A^{(i)}_{\m_1 \dots \m_p\vert \n_1 \dots \n_{p-i}}&=&A^{(i+1)}_{\m_1 \dots \m_p \vert [\n_1 \dots \n_{p-i-1},\n_{p-i}]} 
\nonumber \\
 &+&\, \epsilon(p,i)A^{(i+1)}_{\n_1 \dots \n_{p-i}[ \m_{p-i+1} \dots \m_p \vert \m_1 \dots \m_{p-i-1},\m_{p-i}]} \,,\quad 1\leq i \leq p-1\,, 
\nonumber 
\eqn
where $\epsilon(p,i)\equiv\frac{p!}{(i+1)!(p-i)!}\,$. \\

 \subsection{Standard cohomological results}
The knowledge of $H^{D,0}(s \vert d)$ requires the computation of 
the following cohomological groups: $H(\g)$, $H(\g \vert d)$ in strictly 
positive antighost number, $H(\d \vert d)$ and $H^{inv}(\d \vert d)$.
Some of these are already known \cite{BBH1,BBH2} and will be given in Lemmas, 
whereas some have to be computed, which we do in the following Theorems.

\subsubsection{Cohomology of $\g$}
\label{cohog} 
The cohomology of $\g$ (space of solutions of $\g a
= 0$ modulo trivial coboundaries of the form $\g b$) must be
explicitly worked out and turns out to be isomorphic to the space of functions of 
the following variables:
 \begin{itemize}
 \item  the antifields and all their derivatives, denoted by
 $[\Phi^*]$, \item  the undifferentiated ghost
 $A^{(p)}_{[\m_1 \dots \m_p]}$,
 \item the following undifferentiated ``field strength" of the ghost 
$A^{(p)}$ : $H^A_{[\m_0 \dots \m_p]}\equiv\pa_{[\m_0}A^{(p)}_{\m_1 \dots \m_p]}$, 
\item the curvature $K$ defined in (\ref{fieldstrength}) and all its
 derivatives, denoted by $[K]$.
 \end{itemize}
  The ghost-independent polynomials
 $\a([K],[\Phi^*])$ 
are called ``invariant
 polynomials".

 \vspace{.3cm} \noindent {\bf Comments} \vspace{.1cm}

 Let $\left\{\omega^I\left(A^{(p)}_{[\m_1 \dots \m_p]},H^A_{[\m_0 \dots \m_p]}\right)\right\}$ be
 a basis of the algebra of polynomials in the variables $A^{(p)}_{[\m_1 \dots \m_p]}$
 and $H^A_{[\m_0 \dots \m_p]}$. Any element of $H(\g)$ can be decomposed in
 this basis, hence for any $\g$-cocycle $\a$
 \be\gamma\a=0 \quad\Leftrightarrow\quad
 \a=\a_I([K_{\m_1 \dots \m_{p+1} \vert \n_1 \dots \n_{p+1} }],[\Phi^*])\;
 \omega^I\left(A^{(p)}_{[\m_1 \dots \m_p]},H^A_{[\m_0 \dots \m_p]}\right) + \g \b
 \label{gammaa}\ee where the $\a_I$ are invariant polynomials.
 Furthermore, $\a_I\omega^I$ is $\g$-exact if and only if all the
 coefficients $\a_I$ are zero \be \a_I\omega^I=\gamma\b,\quad
 \Leftrightarrow\quad \a_I=0,\quad\mbox{for
 all}\,\,I.\quad\label{gammab}\ee 
It should also be noted that, as $p >1\,$, there is no ghost in $H(\g)$ with pureghost number 
$1 \leq k < p$, nor is there a ghost in $H(\g)$ with pureghost number $p+1$ 
.

 \subsubsection{General properties of $H(\g \vert d)$}
 The cohomological space $H(\g \vert d)$ is the space of
 equivalence classes of forms $a$ such that $\g a+db=0$, identified
 by the relation $a\sim a'$ $\Leftrightarrow$ $a'=a+\g c+df$. We
 shall need properties of $H(\g \vert d)$ in strictly positive
 antighost (= antifield) number.  To that end, we first recall the
 following result on invariant polynomials (pure ghost number
 $=0$):
 \begin{lemma}\label{2.2}
 In form degree less than n and in antifield number strictly
 greater than $0$, the cohomology of $d$ is trivial in the space of
 invariant polynomials.
 \end{lemma}See \cite{DVHTV}.\\

 Lemma \ref{2.2}, which deals with $d$-closed
 invariant polynomials that involve no ghosts (one considers only
 invariant polynomials), has the following useful consequence on
 general $\g$-mod-$d$-cocycles with $antigh >0\,$, but possibly 
 {\it puregh} $\neq 0\,$.

 \vspace{.3cm} \noindent {\bf{Consequence of Lemma \ref{2.2}}}
 \vspace{.1cm}

 {\it{If $a$ has strictly positive antifield number (and involves
 possibly the ghosts), the equation $\gamma a + d b = 0$ is
 equivalent, up to trivial redefinitions, to $\gamma a = 0.$ That
 is,
 \begin{equation}
 \left.
 \begin{array}{c}
 \gamma a + d b = 0, \\
 antigh(a)>0
 \end{array}
 \right\} \Leftrightarrow \left\{
 \begin{array}{c}
 \gamma a' = 0\,, \\
 a'=a+dc
 \end{array}\right.\,.\label{blip}
 \end{equation}
 Thus, in antighost number $>0$, one can always choose
 representatives of $H(\g \vert d)$ that are strictly annihilated
 by $\g$}}. See \cite{BBH1,BBH2}.

 \subsubsection{Characteristic cohomology $H(\d \vert d)$}
 We now turn to the groups $H(\d \vert d)$, i.e., to the solutions
 of the condition $\d a + db = 0$ modulo trivial solutions of the
 form $a=\d m + dn$. 

 \begin{lemma}
 \label{vanishing} The cohomology groups $H^D_q(\delta \vert d)$
 vanish in antifield  number $q$ strictly greater than $p+1$, \bqn
 H^D_q(\delta \vert d) = 0 \, \hbox{ for } q>p+1. \nonumber\eqn
 \end{lemma}See \cite{BBH1,BBH2}.

 \begin{theorem}
 \label{conservation2} A complete set of representatives of
 $H^D_{p+1}(\d \vert d)$ is given by the antifields $A^{*(p)[\m_1 \dots \m_p]}$,
 {\it{i.e.}}, 
 \bqn 
 \d a^D_{p+1}+d a^{D-1}_{p}&=&0 \nonumber \\
 \Rightarrow a^D_{p+1}&=&\l_{[\m_1 \dots \m_p]}A^{*(p)[\m_1 \dots \m_p]} dx^0\wedge dx^1\wedge\ldots\wedge
 dx^{D-1}+\d b_{p+2}^D+d b_{p+1}^{D-1} 
 \nonumber
 \eqn 
where the $\l_{[\m_1 \dots \m_p]}$ are
 constants.
 \end{theorem}
The proof runs as in the particular case $p=1$ treated in Ref. \cite{BDGH} and will not be repeated here. Representatives with an explicit $x$-dependence were not considered, 
since they would not lead to Poincar\'e-invariant deformations.  

 \begin{theorem}
 \label{conservation3} The cohomology group $H^D_p(\d \vert d)$ vanishes: 
 \bqn \d a^D_p+d a^{D-1}_{p-1}=0  \Rightarrow a^D_{p}= \d   b^{D}_{p+1} + d b^{D-1}_{p}\,. \nonumber\eqn
\end{theorem}

 As far as Theorem \ref{conservation3} is concerned, statements have been made recently
 in the particular case $p=2$ \cite{Biz}, the proof of which is absent and which are actually inaccurate.
 We therefore assume it is not irrelevant to give the complete proof of this theorem.
 
\vspace{.3cm} \noindent {\bf Proof of Theorem \ref{conservation3}: } \vspace{.1cm}
 $H^D_p(\d \vert d) $ is defined by
\begin{eqnarray}
H^D_p(\d \vert d) \cong\{ a^D_p\quad\vert \quad \d a^D_p + db^{D-1}_{p-1}=0\,,\quad
a^D_p \sim a^D_p + \d\m_{p+1}^{D} + d\m_{p}^{D-1}  \}\,.
\nonumber       
\end{eqnarray}
The most general representative is 
\be
a^D_p=A^{*(p-1)}_{\n_1\ldots\n_p\vert\n_{p+1}}\S^{\n_1\ldots\n_p\vert\n_{p+1}}
d^Dx + \m_p^D + \d\m_{p+1}^{D} + d\m_{p}^{D-1}\,,
\label{candidate}
\ee
where $\m_p^D$ is quadratic or more in the antifields.
Acting on $a^D_p$ with the Koszul-Tate differential yields
\begin{eqnarray}
        \d a^D_p = (-)^{p-1}\pa_{\s}{A^{*(p-2)}}^{\m_1\ldots\m_p\vert\m_{p+1}\s}
        \S_{\m_1\ldots\m_p\vert\m_{p+1}}d^Dx + \d\m_p^D - d\d \m_{p}^{D-1}\,.
\end{eqnarray}
Taking the Euler-Lagrange derivative with respect to $A^{*(p-2)\m_1\ldots\m_p\vert\m_{p+1}\s}$ and demanding that $a^D_p$ 
should belong to $H^D_p(\d \vert d)$ gives the weak equation\footnote{A weak equality 
denotes an equality up to terms that vanish on the surface of the solutions of the equations of motion.} 
\begin{eqnarray}
         \mathbf{Y}^{[p,2]}\S_{\m_1\ldots\m_p\vert\n_{1},\,\n_2}\approx 0\,, 
\label{killingweak}
\end{eqnarray}
where $\mathbf{Y^{[p,2]}}$ projects on the Young-tableau symmetry $[p,2]\,$.
Symbolically, the equation (\ref{killingweak}) reads 
\begin{eqnarray}
        d^{\{2\}}\S\approx 0\,,
\end{eqnarray}
where $d^{\{2\}}$ acts on the complex $\Omega_{(2)}(\mathbb{R}^D)$ \cite{BB1}.
Diagrammatically, the action of $d^{\{2\}}$ on the $[p,1]$-type tensor $\S$ looks like
\be 
d^{\{2\}}
\hspace*{2mm}\vspace{.8cm} 
\begin{picture}(30,30)(0,0)
\multiframe(0,10)(10.5,0){1}(10,10){\ft$1$}
\multiframe(10.5,10)(10.5,0){1}(17,10){\ft$p\!\!+\!\!1$}
\multiframe(0,-0.5)(10.5,0){1}(10,10){\ft$2$}
\multiframe(0,-18)(10.5,0){1}(10,17){$ $} \put(4,-13.5){$\vdots$}
\multiframe(0,-28.5)(10.5,0){1}(10,10){\ft$p$}
\end{picture}
\sim\hspace*{2mm} 
\begin{picture}(25,30)(0,0)
\multiframe(0,10)(10.5,0){1}(10,10){\ft$1$}
\multiframe(10.5,10)(10.5,0){1}(17,10){\ft$p\!\!+\!\!1$}
\multiframe(10.5,-0.5)(10.5,0){1}(17,10){\ft$\partial$}
\multiframe(0,-0.5)(10.5,0){1}(10,10){\ft$2$}
\multiframe(0,-18)(10.5,0){1}(10,17){$ $} \put(4,-13.5){$\vdots$}
\multiframe(0,-28.5)(10.5,0){1}(10,10){\ft$p$}\put(34,-1){.}
\end{picture}
\ee 
One defines the action of $d^{\{1\}}$ similarly \cite{BB1}:  
\be 
d^{\{1\}}
\hspace*{2mm}\vspace{.8cm} 
\begin{picture}(30,30)(0,-10)
\multiframe(0,10)(10.5,0){1}(10,10){\ft$1$}
\multiframe(10.5,10)(10.5,0){1}(10,10){}
\multiframe(0,-0.5)(10.5,0){1}(10,10){\ft$2$}
\multiframe(0,-18)(10.5,0){1}(10,17){$ $} \put(4,-13.5){$\vdots$}
\multiframe(0,-28.5)(10.5,0){1}(10,10){\ft$p$}
\end{picture}
\sim\hspace*{2mm} 
\begin{picture}(25,30)(0,-10)
\multiframe(0,10)(10.5,0){1}(10,10){\ft$1$}
\multiframe(10.5,10)(10.5,0){1}(10,10){}
\multiframe(0,-0.5)(10.5,0){1}(10,10){\ft$2$}
\multiframe(0,-18)(10.5,0){1}(10,17){$ $} \put(4,-13.5){$\vdots$}
\multiframe(0,-28.5)(10.5,0){1}(10,10){\ft$p$}
\multiframe(0,-39)(10.5,0){1}(10,10){\ft$\partial$}
\put(34,-11){.}
\end{picture}
\ee 
On the other hand, the differentials $d_1$ and $d_2$ on a $[p,q]$-type 
tensor $T$ are defined as follows :  
\begin{eqnarray}
        (d_1T)_{\a\m_1\ldots\m_p\vert\n_1\ldots\n_q}&\equiv& 
        \pa_{[\a}T_{\m_1\ldots\m_p]\vert\n_1\ldots\n_q}\,,
        \nonumber \\
        (d_2T)_{\m_1\ldots\m_p\vert\n_1\ldots\n_q \a}&\equiv& 
        T_{\m_1\ldots\m_p\vert[\n_1\ldots\n_q,\a]}\,.
\end{eqnarray}
Operating with $d^{\{1\}}$ on equation (\ref{killingweak}) we have 
        $d^{\{1\}}d^{\{2\}}\S\approx 0$ or $d_1 d_2\S\approx 0\,$, since 
        $d^{\{1\}}d^{\{2\}}\equiv d_1d_2$ in $\Omega_{(2)}(\mathbb{R}^D)$ \cite{BB1}.
Considering the second column of the latter equation and using the isomorphism \cite{BBH1,BBH2} $H_0^1(d\vert \d)\cong H^D_{D-1}(\d\vert d)\cong 0$ where the last 
isomorphism holds because $D>p+2\,$, we have, in component : 
        $\pa_{[\r}\S_{\m_1\ldots\m_p]\vert\n_1}\approx \pa_{\n_1}B_{\m_1\ldots\m_p\r}\,$, 
  where $B_{\m_1\ldots\m_p\r}$ is a tensor completely antisymmetric in its $p+1$ 
  indices.
 Antisymmetrizing the above equation in all its indices gives 
        $0\approx \pa_{[\n}B_{\m_1\ldots\m_p\r]}$
 which defines a representative of the group $H^{p+1}_0(d\vert\d)$, itself  
 isomorphic to $H^{D}_{D-p-1}(\d\vert d)\,$. The latter group vanishes for $D>2p+2$ and 
 is given by constants when $D=2p+2\,$.
 Consequently, one has 
 $B_{\m_1\ldots\m_p\r}\approx D_{\m_1\ldots\m_p\r}\d^D_{2p+2}+\pa_{[\r}C_{\m_1\ldots\m_p]}$ and 
$\pa_{[\r}\S_{\m_1\ldots\m_p]\vert\n}\approx \pa_{\n}\pa_{[\r}C_{\m_1\ldots\m_p]}\,$.
Solving this equation using $H^{p}_{0}(d\vert \d)\cong H^{D}_{D-p}(\d\vert d)\cong 0$ 
for $D>2p+1$ gives 
$ \S_{\m_1\ldots\m_p\vert \n}-C_{\m_1\ldots\m_p ,\n}\approx $  
$\pa_{[\m_1}M_{\m_2\ldots\m_p]\vert\n}$ for a given $M_{\m_2\ldots\m_p\vert\n}$ completely 
antisymmetric in its first $p$ indices. Its irreducible decomposition under $GL(D)$ 
consists in a sum of two types of fields :  
$M_{\m_2\ldots\m_p\vert\n} \equiv M^A_{\m_2\ldots\m_p\n} + M^B_{\m_2\ldots\m_p\vert\n}$  
where $M^A$ is completely antisymmetric and $M^B$ is of symmetry type $[p,1]\,$. 
 Antisymmetrizing the above equation relating $\S, C$ and $M$ yields ($\S$ and $M^B$ 
 both vanish due to their symmetry properties) 
 $-C_{[\m_1\ldots\m_p,\n]}\approx \pa_{[\m_1}M^A_{\m_2\ldots\m_p\n]}\,$. The latter 
 equation, once solved, gives 
 $M^A_{\m_2\ldots\m_p\m_1}\approx C_{\m_1\ldots\m_p} +\pa_{[\m_1}S_{\m_2\ldots\m_p]}$. 
 The tensor $\S$ then decomposes as 
\begin{eqnarray}
        \S_{\m_1\ldots\m_p\vert \n}\approx C_{\m_1\ldots\m_p,\n}+
        \pa_{[\m_1}M^B_{\m_2\ldots\m_p]\vert\n}+C_{\n[\m_2\ldots\m_p,\m_1]}\,,
\end{eqnarray}where $S$ has been eliminated by redefining $M^B\,$.
Substituting this decomposition back into (\ref{killingweak}) brings constraints 
on $C$ and $M^B\,$. After some algebra, one finds  
$\pa_{[\n_1}M^B_{\n_2\ldots\n_p]\vert[\n_{p+1},\n_{p+2}]}\approx 0\,$. 
Again, considering the second column and using the isomorphism 
$H_0^1(d\vert \d)\cong H^D_{D-1}(\d\vert d)\cong 0$ (remember that $D>2p+2$), 
we write $\pa_{[\n_1}M^B_{\n_2\ldots\n_p]\vert\n_{p+1}}\approx \pa_{\n_{p+1}}B_{[\n_1\ldots\n_p]}\,$. Antisymmetrizing the latter equation and 
using  $H_0^p(d\vert \d)\cong H^D_{D-p}(\d\vert d)\cong 0$ ($D>2p+1$), we 
obtain $\pa_{[\n_1}M^B_{\n_2\ldots\n_p]\vert\n_{p+1}}\approx$
$\pa_{\n_{p+1}}\pa_{[\n_1}N_{\n_2\ldots\n_{p}]}\,$.
Defining $\tilde{C}_{\n_1\ldots\n_{p}}=C_{\n_1\ldots\n_{p}}+\pa_{[\n_1}N_{\n_2\ldots\n_{p}]}\,$ 
we finally get $\S_{\m_1\ldots\m_{p}}\approx \tilde{C}_{\m_1\ldots\m_{p},\n}+$
$\tilde{C}_{\n[\m_2\ldots\m_{p},\m_1]}\,$. 

As a result, $a^D_p$ in Eq. (\ref{candidate}) reads 
\begin{eqnarray}
        a^D_p &=& (\tilde{C}_{\m_1\ldots\m_{p},\n}+\tilde{C}_{\n\m_2\ldots\m_{p},\m_1})
A^{*(p-1)\m_1\ldots\m_p\vert \n}d^Dx+\m^D_p+\d\m_{p+1}^{D} + d\m_{p}^{D-1}
\nonumber \\
&=&\frac{(p+1)}{p}\tilde{C}_{\m_1\ldots\m_{p},\n}A^{*(p-1)\m_1\ldots\m_p\vert \n}d^Dx+\m^D_p+\d\m_{p+1}^{D} + d\m_{p}^{D-1}
\nonumber \\
&=&-\frac{(p+1)}{p}\tilde{C}_{\m_1\ldots\m_{p}}\pa_{\n}A^{*(p-1)\m_1\ldots\m_p\vert \n}d^Dx+\m^D_p+\d\m_{p+1}^{D} + d{\m'}_{p}^{D-1}\,.
\end{eqnarray}
Since $\d A^{*(p)\m_1 \dots \m_p}=(-)^p \pa_{\n}A^{*(p-1)\m_1 \dots \m_p\vert \n}\,$, 
we see that only non-trivial part of $a^D_p$ must be in the term $\m^D_p$ which is 
quadratic or more in the antifields. However, such representatives of $H^D_p(\d\vert d)$ are trivial \cite{BBH1},  which implies $H^D_p(\d\vert d)\cong 0$ in the theories under 
consideration. This proves Theorem \ref{conservation3}.
 
 \subsubsection{Invariant characteristic cohomology : $H^{inv}(\d \vert d)$} 
 
 The crucial result that underlies all consistent interactions
 does not deal with the general cohomology of $\delta$ modulo $d$ but
 rather with the {\em invariant} cohomology of $\delta$ modulo $d$.
 The group $H^{inv}(\delta \vert d)$ is important because it
 controls the obstructions to removing the antifields from a
 $s$-cocycle modulo $d$ \cite{BBH1,BBH2}.

 The central theorem that gives $H^{inv}(\delta \vert d)$ in
 antighost number $\geq p$ is
 \begin{theorem}\label{2.6}
 Assume that the invariant polynomial $a_{k}^{D}$
 ($k =$ antifield number)
 is $\delta$-trivial
 modulo $d$,
 \be
 a_{k}^{D} = \delta \mu_{k+1}^{D} + d \mu_{k}^{D-1} ~ ~ (k \geq p).
 \label{2.37}
 \ee
 Then, one can always choose $\mu_{k+1}^{D}$ and $\mu_{k}^{D-1}$ to be
 invariant.
 \end{theorem}
The proof of this theorem proceeds exactly as the proofs of
 similar theorems established for e.g. vector fields \cite{BBH2} or 
 gravity \cite{BDGH}. Notice however that there is a subtlety one should not overlook in the 
case $k=p\,$; the Appendix is devoted to clarify this specific issue in the particular 
case $p=2\,$ (to fix the ideas). 
For $p>2\,$, the proof follows exactly the same lines. 
\vspace*{.2cm}

 As a consequence of Theorem \ref{2.6}, we have $H^{D,inv}_k(\delta \vert d) = 0$ for $k>p+1\,$, 
 while $H^{D,inv}_{p+1}(\delta \vert d)=H^{D}_{p+1}(\delta \vert d)$ is given by Theorem
 \ref{conservation2} and $H^{D,inv}_p(\delta \vert d)$ vanishes, by Theorem
 \ref{conservation3}.

 \subsection{First order consistent interactions}

 We can now proceed with the derivation of
 the cohomology of $s$ modulo $d$ in form degree $D$ and in ghost
 number zero. A cocycle $a$ of $H^{0,D}(s \vert d)$ must obey 
 \be 
 s a + d b = 0\,.
 \label{cocycsd} 
 \ee 
 To analyze (\ref{cocycsd}), we expand
 $a$ and $b$ according to the antifield number, $a = a_0 + a_1 +
 ...+a_k $, $b = b_0 + b_1 + ...+b_k  $, where locality implies that the expansion 
 stops at some finite antifield number \cite{BBH1, BBH2}. We recall \cite{BH93}
 (i) that the antifield-independent piece $a_0$ is the deformation
 of the Lagrangian; (ii) that $a_1$ contains the informations about the
 deformation of the gauge transformations;  
(iii) that $a_2$ contains the 
 informations about the deformation of the gauge algebra and  
 of the first-stage reducibility conditions; and (iv) that the $a_k$ ($k\geq 3$) give the
 informations about the higher-stage reducibility conditions and the deformation of the higher order structure
 functions, which appear only when the algebra does not close
 off-shell.  Thus, if one can show that the most general solution
 $a$ of (\ref{cocycsd}) stops at $a_0$, one can conclude that the gauge algebra
  is rigid and that the gauge transformations are not deformed at first order. 

 Writing $s$ as the sum of $\gamma$ and $\delta$, the equation $s a
 + d b = 0$ is equivalent to the system of equations $\delta a_i +
 \gamma a_{i-1} + db_{i-1} = 0$ for $i = 1, \cdots, k$, and $\gamma
 a_k + db_k =0$.

 \subsubsection{Terms $a_k$, $k>p+1$}

Following a general procedure used e.g. in \cite{BDGH}, one can show that
the  terms $a_k$ ($k>p+1$) may be discarded one after another from the aforementioned descent.
The latter terminates thus with $\g a_{p+1} + d b_{p+1} =0\,$. Using the consequence of Lemma \ref{2.2}, this equation is equivalent to $\g a_{p+1}=0\,$.

 Note that this result is independent of any condition on the number of
 derivatives or of Poincar\'e invariance.  These requirements have not
 been used so far. The crucial ingredient of the proof is that the
 cohomological groups $H^{inv}_k(\delta \vert d)$, which control
 the obstructions to remove $a_k$ from $a$, vanish for $k>p+1$.

 \subsubsection{Computation of $a_{p+1}$}

We thus have the following descent:
 \bqn
 \d a_1 + \g a_0 +d b_0 &=&0~,
 \label{5.4}
 \\
 &\vdots&
 \nonumber \\
 \d a_{p+1} + \g a_p +d b_p &=&0~,
 \label{5.6}
 \\
 \g a_{p+1}&=&0~. \label{5.7} \eqn 
The last equation implies $a_{p+1}=\a_I\o^I\,$.
Acting with $\g$ on the second to last equation and using Lemma \ref{2.2} leads to $b_p=\b_I\o^I$.
 Substituting those expressions in Eq. (\ref{5.6}), we find that a necessary (but not sufficient) condition for $a_{p+1}$ to be a non-trivial solution of (\ref{5.6}),
 such that $a_p$ exists, is that $\a_I$ be a non-trivial element of
 $H^D_{p+1}(\d\vert d)$. Asking for Poincar\'e invariance, Theorem \ref{conservation2} then imposes
 $\a_I \sim A^{*(p)[\m_1 \dots \m_p]}$. We next have to complete this $\a_I$ with an
 $\o^I$ of ghost number $p+1$ in order to build a ghost zero candidate
 $\a_I\o^I$ for $a_{p+1}$. However, there is no $\o^I$ with pureghost number $p+1$, thus $a_{p+1}$ must be zero.

 \subsubsection{Computation of $a_k$, $0<k<p+1$}
 
 Following the same steps as for $a_{p+1}$, $a_p$ writes $ a_{p}=\a_I\o^I\,$, 
 where $\a_I$ belongs to $H^D_p(\d\vert d)$. As shown in Theorem \ref{conservation3}, 
 this group vanishes. Using Theorem \ref{2.6}, it can be shown that one is allowed to set $a_p=0$ (see \cite{BH93}).\\
Continuing the analysis, we have to find $a_{p-1}$ of the form $a_{p-1}=\a_I\o^I\,$, where
$\a_I$ is a non-trivial element of
 $H^D_{p-1}(\d\vert d)$. The ghost number of $\o^I$ must be $p-1$, but there is no such $\o^I$,
so $a_{p-1}$ vanishes.
One can repeat this argument for antifield number $k-2$, etc. until one reaches antifield number 0, where the argument does not work anymore.

At this point, we have proved the rigidity of the gauge algebra and gauge transformations. This has been done under the sole assumptions of locality and Poincar\'e invariance (besides the smoothness in the deformation parameter).

 \subsubsection{Computation of $a_0$}
 
 We are now reduced to solve the equation $\g a_0 +d b_0=0$ for $a_0\,$.
 Such an $a_0$ corresponds to deformations of the Lagrangian which are
 gauge invariant up to a total derivative. The Euler-Lagrange derivatives $\frac{\d a_0}{\d C}$ must be gauge invariant and must satisfy Bianchi identities of the type  
 (\ref{Bianchi}) (because of the gauge invariance of $\int a_0$).
 Asking that $a_0$ should not contain more than two derivatives, we obtain that 
 $\frac{\d a_0}{\d C}$ must be at most linear in the curvature $K\,$. These three conditions together completely constrain $a_0\,$ and have only two solutions.
 The first one is a cosmological-constant-like term  
 \begin{eqnarray}
a_0=\Lambda \h_{\m_1 \n_1} \ldots \h_{\m_p \n_p}C^{\m_1 \ldots \m_p \vert \n_1 \ldots \n_p}\,.
\end{eqnarray}
 The second one, where $\frac{\d a_0}{\d C}$ are linear in the curvature $K$,
 is the free Lagrangian itself  \cite{dMH2}. 
 
 So we conclude that, apart from a cosmological-constant-like term, the deformation only changes the coefficient of the free Lagrangian and is not essential.  
 
 \section{Comments and conclusions}
 \setcounter{equation}{0} \setcounter{theorem}{0}
 \setcounter{lemma}{0}

 We can summarize our results as follows :  under the hypothesis of
locality and Poincar\'e invariance,
 there is no smooth deformation of the free theory which modifies
 the gauge algebra or the gauge transformations. If one further excludes deformations involving more than two derivatives in the Lagrangian, then the only smooth deformation of the free theory is a cosmological-constant-like term.

 Without this extra condition on the derivative order, one can
 introduce Born-Infeld-like interactions that involve powers of the
 gauge-invariant curvatures $K\,$. Such
 deformations modify neither the gauge algebra nor the gauge
 transformations.

 We believe to have reached a fairly high degree of generality, these results being obtained 
under very few assumptions and constraining a whole class of theories for the gauge 
fields $C_{\m_1\ldots\m_p\vert\n_1\ldots\n_p}\,$, with $p\,$ greater than one but 
 otherwise arbitrary. The next step would be to consider arbitrary exotic ``spin-2'' fields 
represented by Young tableaux having two columns of different arbitrary lengths. 
It is planned to return to this question in the future.

 \section*{Acknowledgments}

We thank X. Bekaert for his comments. 
S.C. is grateful to M. Henneaux for enlightening discussions. 
The work of N.B. is supported by a Wiener-Anspach fellowship (Belgium), 
while the work of S.C. is supported in part by the ``Actions de Recherche Concert{\'e}es'' of
the ``Direction de la Recherche Scientifique - Communaut{\'e}
Fran{\c c}aise de Belgique'', by a ``P\^ole d'Attraction
Interuniversitaire'' (Belgium), by IISN-Belgium (convention
4.4505.86) and
by the European Commission RTN program HPRN-CT-00131, in which
she is associated to K. U. Leuven.

\appendix
\section{Appendix}

At some stage of the proof of Theorem \ref{2.6} for $k=p=2\,$ (here we consider $p=2$ to fix 
the ideas, but the general $p$-arbitrary case proceeds exactly in the same way), 
one needs the following Theorem:
\begin{theorem}\label{projection}
 Let the zero-antighost-number invariant polynomial $\a_0$ satisfy 
\be 
\alpha_{0 \, \m \n \vert  \r} = \delta Z_{1\, \m \n \vert \r} + 
(\pa_{\r} W_{0\, \m \n} - \pa_{[\r} W_{0\, \m \n]})\,, 
\label{proj1}
\ee 
where $\a_0$ and $Z_1$ are [2,1]-tensors and $W_0$ is a [2,0]-tensor, then
\bqn Z_{1\, \m \n \vert \r} &=  &Z'_{1\, \m \n \vert \r}+\d \phi_{2\, \m \n \vert \r} 
+ (\pa_{\r} \chi_{1 \, [ \m \n]}-\pa_{[\r} \chi_{1 \, \m \n]})\,,\\
W_{0\, \m \n}&= &W'_{0\, [\m \n]} + \d  \chi_{1\, [\m \n]}\,,
\eqn
for some invariants $Z'_{1}$ and $W'_{0}$.
\end{theorem}
To prove this Theorem, we need the following Lemma:
\begin{lemma} 
\label{lemd}
Suppose that Theorem \ref{2.6} is proved for $k>p\,$.
If $\a_0^1$ is an invariant polynomial and satisfies 
\be \alpha_{0 }^1 = \delta Z_{1}^1 + d W_{0}^0 \,, 
\label{proj1bis}
\ee 
then, for some invariant polynomials ${Z'}_{1}^1$ and ${W'}_{0}^0\,$,
\bqn Z_1^1 &=  &{Z'}_1^1+\d \phi_2^1 + d \chi_{1}^0\,, \label{une}\\
W_{0}^0&= &{W'}_{0}^0 + \d  \chi_1^0\,. \label{deux}
\eqn
\end{lemma}

Though Lemma \ref{lemd} is already known \cite{Bernard,BBH}, 
the proof does never appear explicitely. We shall therefore give it here.\\
Using standard techniques, one gets the following descent
\bqn
\a_1^2&= &\d Z_2^2 + d Z_1^1 \label{prems}\\
&\vdots &\nonumber \\
\a_{D-1}^D &= &\d Z_D^D + d Z_{D-1}^{D-1}\,, 
\nonumber  
\eqn
where all the $\a_{i-1}^i$ are invariant. 
As $D-1 \geq p+1$, by Theorem \ref{2.6} we can choose $Z_D^D$ and $Z_{D-1}^{D-1}$ invariant. 
The invariance property propagates up until
$\a_1^2= \d {Z'}_2^2 + d {Z'}_1^1 $, where ${Z'}_2^2$ and ${Z'}_1^1$ have been chosen 
invariant.  
Substracting the latter equation from (\ref{prems}) and knowing that 
$H_1^1(\d\vert d) \cong H_D^D(\d\vert d)$ vanishes, we get (\ref{une}). 
Substituting  (\ref{une}) in (\ref{proj1bis}) and acting with $\g$, we find 
$d ( \g (W^0_0-\d \chi_{1}^0)) = 0\,$. Using the algebraic Poincar\'e lemma and the fact that 
there is no constant with positive pureghost number, this implies $\g (W^0_0-\d \chi_{1}^0)=0\,$, 
which in turn gives (\ref{deux}), as there exists no $\g$-exact term of pureghost number $0\,$.
\vspace*{.5cm}

\noindent {\bf Proof of Theorem \ref{projection}:}
The first step is to constrain the last term of (\ref{proj1}).
In (hyper)form notation, Eq. (\ref{proj1}) reads:
\be 
\a_0^{[2,1]}= \delta Z_1^{[2,1]}+ d^{\{2\}}W_0^{[2,0]}\,. \label{proj}
\ee
Acting with $ d^{\{1\}}$ on Eq. (\ref{proj}), we get
\be  
d^{\{1\}}\a_0^{[2,1]}=\delta (d^{\{1\}} Z_1^{[2,1]})+ d^{\{2\}} (d^{\{1\}}W_0^{[2,0]})\,. 
\label{last}
\ee
We now consider only the second column. 
Since $d^{\{2\}}d^{\{1\}}W_0^{[2,0]}\equiv d_2d_1W_0^{[2,0]}\,$, 
Equation (\ref{last}) reads 
\be 
\tilde{\a}_{0}^1 = \d \tilde{Z}_1^1 + d \tilde{W}_0^0\,. 
\ee
As $\tilde{\a}_{0}^1$ is invariant, we know by Lemma \ref{lemd} that 
$\tilde{W}_0^0 = \tilde{W}_{~0}^{'0} + \d \b_1^0\,$, where $\tilde{W}_{~0}^{'0}$ is invariant.
Remembering that $\tilde{W}_0^0 =d_1W_0^{[2,0]}\,$, we have in components:
$\pa_{[\s} W_{0\, \m \n]}= {\tilde{W}'}_{0\, [\m \n \s]} + \d \b_{1\, [\m \n\s]}$ for some 
invariant ${\tilde{W}'}_{0\, [\m \n\r]} \,$.
Plugging this in (\ref{proj1}), we get
\be 
\alpha_{0 \, \m \n \vert  \r} +{\tilde{W}'}_{0\, [\m \n \r]}= \delta (Z_{1\, \m \n \vert \r} -  
\b_{1\, [\m \n\r]})+ \pa_{\r} W_{0\, \m \n}\,. 
\ee
Considering only $\r$ as a form index, we have 
\be \bar{\a}_{0 \vert [\m \n]}^1=  \d \bar{Z}_{1\vert [\m \n]}^1 + d W_{0 \vert [\m \n]}^0 \,,\ee
which, as $\bar{\a}_{0 \vert \m \n}^1$ is invariant,  implies that 
$\bar{Z}_{1\vert \m \n}^1 = {\bar{Z}}^{'1}_{~1\vert \m \n} + \delta \chi_{2\vert \m \n}^1 
+d\chi_{1\vert \m \n}^0$ and $W _{0 \vert \m \n}^0 = {W'}^0_{~0 \vert \m \n} 
+ \d  \chi_{1\vert \m \n}^0$, for some invariants ${\bar{Z}}^{'1}_{~1\vert \m \n}$ and 
${W'}^0_{~0 \vert \m \n}$. Equivalently, we have 
\bqn
Z_{1\, \m \n \vert \r} &= &  \b_{1\, [\m \n\r]}+{\bar{Z}'}_{1 \r \vert \vert [\m \n]}+
\d \chi_{2\, \r \vert \vert [\m \n]} + \pa_{\r} \chi_{1 \, [ \m \n]}\,,\label{Z}\\
W_{0\, \m \n}&= &{W'}_{0\, [\m \n]} + \d  \chi_{1\, [\m \n]}.
 \eqn
Removing the completely antisymmetric part from (\ref{Z}), we get the wanted result
\be Z_{1\, \m \n \vert \r} =  {Z'}_{1\, \m \n \vert \r}+\d \phi_{2\, \m \n \vert \r} 
+ (\pa_{\r} \chi_{1 \, [ \m \n]}-\pa_{[\r} \chi_{1 \, \m \n]})\,,\ee
where ${Z'}_{1\, \m \n \vert \r}={\bar{Z}}'_{1 \r \vert [\m \n]}
-{\bar{Z}}'_{1 [\r  \vert \m \n]}$ and 
$\phi_{2\, \m \n \vert \r} = \chi_{2\, \r  \vert [\m \n]}- \chi_{2\, [\r  \vert \m \n]}$ 
are $[2,1]$-tensors, and $Z'_1$ is invariant.
\vspace*{.3cm}

We stress again that the proof in the case where $p$ is arbitrary proceeds exactly in the same way.

 \end{document}